# Corrections on energy spectrum and scatterings for fast neutron radiography at NECTAR facility*

LIU Shu-Quan(刘树全)[1,2], Bücherl Thomas[2], LI Hang(李航)[1,3], ZOU Yu-Bin(邹宇斌)[1], LU Yuan-Rong(陆元荣)[1;1)], GUO Zhi-Yu(郭之虞)[1]

[1] State Key Lab of Nuclear Physics and Technology & School of Physics, Peking University, Beijing 100871, China

[2] Technische Universität München, ZTWB Radiochemie München RCM, Walther-Meissner-Str. 3, Garching 85748, Germany

[3] Institute of Nuclear Physics and Chemistry, Chinese Academy of Engineering Physics, Mianyang 621900, China

## Abstract

Neutron spectrum and scattered neutrons caused distortions are major problems in fast neutron radiography and should be considered for improving the image quality. This paper puts emphasis on the removal of these image distortions and deviations for fast neutron radiography performed at the NECTAR facility of the research reactor FRM-II in Technische Universität München (TUM), Germany. The NECTAR energy spectrum is analyzed and established to modify the influence caused by neutron spectrum, as well as the Point Scattered Function (PScF) simulated by the Monte-Carlo program MCNPX is used to evaluate scattering effects from the object and improve images qualities. Good analysis results prove the sounded effects of above two corrections.

*Supported by National Natural Science Foundation of China under Grant No. 11079001
1) E-mail: yrlu@pku.edu.cn

# 1. Introduction

Fast neutron radiography is a suited detection tool for bulk objects due to the low cross-section of fast neutrons and the elements. For quantitative analysis, such as the defect evaluations and fast neutron tomography, some significant effects should be considered and corrected [1-3].

## 1.1 Problems and methods

In traditional radiography analysis, relationship between the gray values on the image and the sample thickness $t$ is treated by the exponential attenuation law:

$$\frac{I}{I_0} = e^{\Sigma_{tot}(E) \cdot t} \quad (1)$$

where $\Sigma_{tot}(E)$ is the linear attenuation coefficient of the sample as a function of neutron energy $E$, $I_0$ and $I$ is the neutron flux before and after the penetration. Usually a constant $\Sigma_{tot}$ will be used to evaluate the sample thickness, but this will bring in deviations when the neutron beams are polychromatic. This is called the energy spectrum effects in fast neutron radiography [4].

The scattering effects, on the other hand, are mainly caused by the interacted neutrons scattered out from the sample, which make overlaps and blurring on the images. Neutrons recorded on the detectors can be represented as the sum of penetrated neutrons and scattered neutrons $I_s$. Our goal is to evaluate and remove $I_s$ from the original images $I$. Fig. 1 shows the ratios of scattering cross-section to total cross-section ($\sigma_s/\sigma_t$) for different materials at neutron energies between 1.0 and 10 MeV. It is noticed that scatterings effects are more significantly in fast neutron radiography, as can be seen in Figure 1 that more than 90% of the attenuated neutrons are scattered during fast neutron radiography.

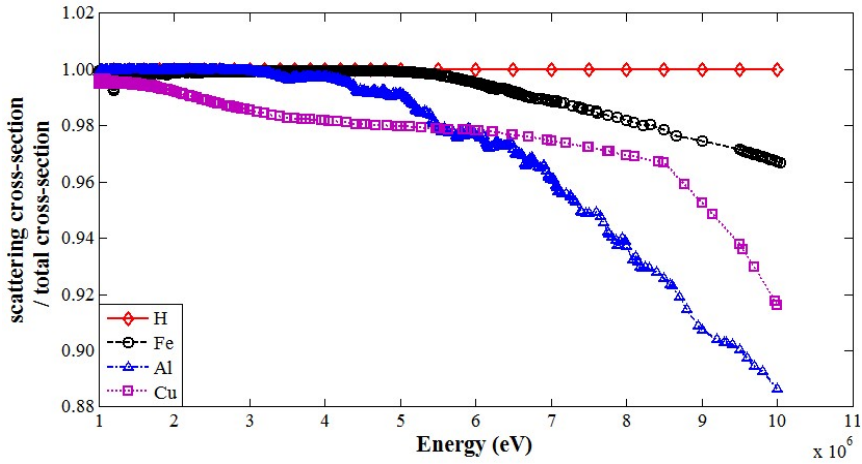

**Fig.1.** Scattering to total cross-section ratios for different elements at fast neutron energies [5]

Considering both energy spectrum and scattering effects, Equation (1) is now fixed as

$$\frac{I - I_s}{I_0} = \frac{\int_0^{E_0} \phi(E) e^{-\Sigma_{tot}(E)t} dE}{\int_0^{E_0} \phi(E) dE} \qquad (2)$$

where $\phi(E)$ is the incident neutron spectrum and $I_s$ is the scattered neutron flux detected.

The evaluation of $I_s$ in fast neutron radiography is a key problem in corrections. Decades ago the way for evaluating $I_s$ is to model the fast neutron experiment conditions and estimate the scattering component with an analytical representation [6-9]. Recently an effective method for scattering correction in thermal neutron radiography is using the Point Scattered Function [10,11]. However, due to the large-size object used in fast neutron radiography, the simulation of PScFs for fast neutrons is time costing and requires high CPU performances. In this work, to solve this problem, Linux cluster and parallel computing technology is applied and PScF method is extent to fast neutrons scattering analysis.

These two effects are studied separately by adjusting the samples in this work. The fast neutron spectrum effects were numerically calculated and applied to the image analysis. PScFs of fast neutrons were simulated and integrated by all beams penetrating the sample to obtain $I_s$ and then it was subtracted from the original image. The quantitative results with and without each effect will be discussed and compared finally.

**1.2 NECTAR facility**

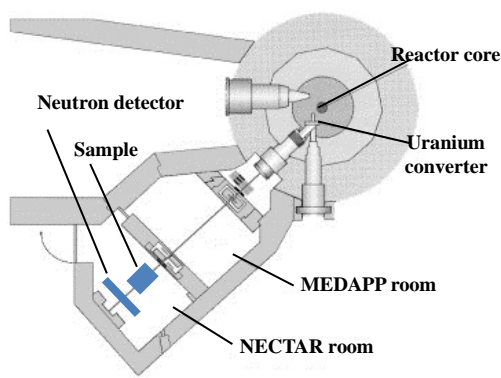

**Fig.2.** Sketch of the fast neutron radiography experiment conditions at NECTAR[12]

This work was performed at the NECTAR (NEutron Computerized Tomography And Radiography) facility located at the research reactor FRM-II in Technische Universität München (TUM), Germany [12]. The layouts of NECTAR and experiment conditions are shown in Figure 2. The source-to-detector distance was $L =$ 9355mm with a neutron beam of divergence 2.1 degree and average neutron energy of 1.9MeV. The neutron flux at sample position is $5.4\times10^5$ $cm^{-2}s^{-1}$ and the $L/D$ ratio is 233. As $L$ and $L/D$ are so large here, parallel beam geometries were used in later calculations. A 300×300×2.4 $mm^3$ pp-converter (doped with 30% ZnS) with a spatial resolution of 0.6mm was used as the detector [13]. The sample-to-detector distance was flexible depends on the experimental requirements.

## 2. Energy spectrum effects

### 2.1 Analysis

Considering the neutron energy spectrum of NECTAR, the expression of $\ln(I/I_0)$ and $t$ treated from Equation (2) is now:

$$-\ln(\frac{I}{I_0}) = -\ln\left(\frac{\int_0^{E_0} \phi(E)e^{-\Sigma_{tot}(E)t}dE}{\int_0^{E_0} \phi(E)dE}\right) \quad (3)$$

For monochromatic neutrons the linear attenuation coefficient $\Sigma tot(E)$ is independent of energy $E$ thus $\ln(I/I_0)$ have a linear relationship with thickness $t$. For polychromatic neutrons, the right side of equation (3) was numerically integrated and fitted by polynomial representations.

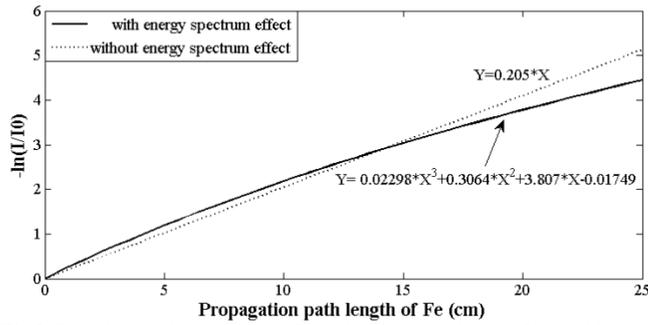

**Fig.3.** The relationship between $ln(I/I_0)$ and penetrating depth $t$ of iron with and without the NECTAR energy spectrum effect. Polynomial fittings are also shown in the figure.

Figure 3 shows the relationships of $-ln(I/I_0)$ and penetrating depth $t$ with and without the energy spectrum effects. It can be seen that below 14cm thickness, the polychromatic neutron beam attenuates faster than monochromatic neutrons. This is due to the thermal neutron components in the beam at NECTAR, which has larger cross-sections with irons and attenuates faster. However, when $t$ is larger than 14cm, less thermal neutrons and relatively more fast neutrons remain in the beams, which makes the beam attenuates slowly.

**2.2 Experiment and results**

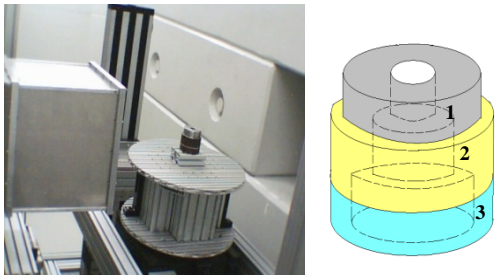

**Fig.4.** Sample group 1. Left: Experiment arrangement for sample group 1 in NECTAR platform; Right: Geometry diagram of sample group 1

A group of three iron cylinders with air drillings was used to check the neutron energy spectrum effects at NECTAR (Fig. 4). The outer and inner diameters of cylinder 1, 2, 3 are: 7.0(2.0), 8.0(3.2) and 8.0(6.2) cm, respectively. To avoid other effects, iron was selected because iron has a relatively low scattering cross-section for fast neutrons compared to other materials. The distance of the center of sample group 1 to detector was set to 380mm so that the sample scatterings can be ignored.

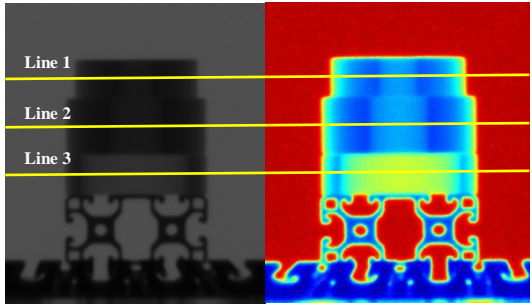

**Fig.5.** Image of sample group 1and data analysis lines. Left: radiographic image. Right: normalized image

The raw image was filtered, processed by open beam image, dark image and finally normalized, as shown in Figure 5. Three lines were drawn crossing the three cylinders and the line data were obtained as three arrays for quantitative analysis.

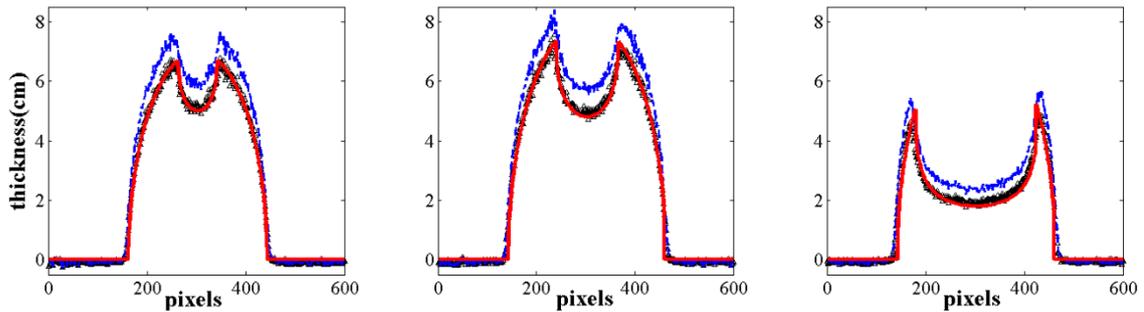

**Fig.6.** Penetration thickness profiles at different positions (a) line1. (b) line 2. (c) line 3. Blue dashed line: Before energy spectrum correction. Black dotted line: After energy spectrum correction. Red solid line: Real sample thicknesses.

Figure 6 shows the penetration thicknesses derived from Equation (3) based on the normalized images in Figure 5. It can be seen that there will be a 10-20% deviation if simply using the exponential law. After considering the energy spectrum, the deviation was eliminated and the results show a precise thickness profiles as the real sample thickness.

## 3. Neutron scattering effects

### 3.1 Point Scattered Functions (PScF)

Unlike the PSF (Point Spread Function) that used to describe the whole

projection of a point in the specimen, PScF (Point Scattered Functions) is used to describe the density distribution of scattered neutrons on the detector from a point neutron source hitting directly to the material [14]. In brief, PScF records all the scattered neutrons that are not along the beam line, as shown in Figure 7.

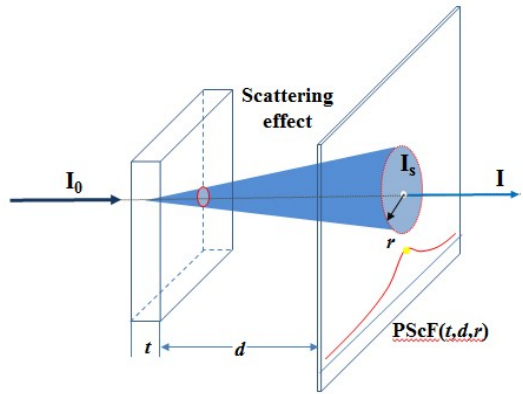

**Fig.7.** The diagram of PScF of a neutron beam as a function of *t,d,r*

The PScFs of fast neutrons were simulated by the Monte Carlo program MCNPX on Linux cluster at the Leibniz Rechenzentrum [15]. To obtain the PScF data for a 10cm thickness polyethylene with an error less than 5%, a number of $10^8$ neutrons were simulated, which takes about 10 hours with 64 CPUs parallel computing.

The simulation parameters were set as follows. The geometry arrangement of NECTAR was based on the work of Harald Bretkreutz in his Ph.D diploma thesis [16]. The neutron source is carried out as a 15cm×15cm area source with each point an isotropic divergence point source. The energy spectrum of each point source was characterized by the neutron spectra in NECTAR with energy range from 0.001eV to 14MeV [12]. As a simplification, we ignore the filters in the geometry layout. The fast neutrons are collimated and scattered by the architectures and finally arrived at NECTAR. The detector was described by a 40cm-radius mesh tally with a series of rings. Each ring has a radius interval of 0.1mm. Neutrons scattered into the rings will be counted and converted to the PScF density distributions on the detector. Different sample materials, thickness $t$ and sample-to-detector distance $d$ were set and simulated.

Figure 8 and Figure 9 show the simulated PScFs for polyethylene with different sample thicknesses $t$ and sample-to-detector distances $d$. It can be seen from Figure 8 that with the increase of $d$, the scattering effect becomes weaker. Analytically, $S(d,t)$ has an inverse proportional relationship with the square of $d$ for thermal neutrons[10]. Scattering effects can be neglected when the distance is larger than 100mm. Unfortunately, the increase of $d$ would increase the geometric sharpness and reduce the resolution of radiography images. On the other hand, Figure 9 shows that when the sample thickness goes thicker, the scatterings effects will increase firstly and then decrease. This can be explained that in thicker samples many scattered neutrons are multiply scattered and vanished inside of the sample instead of escaping and being detected.

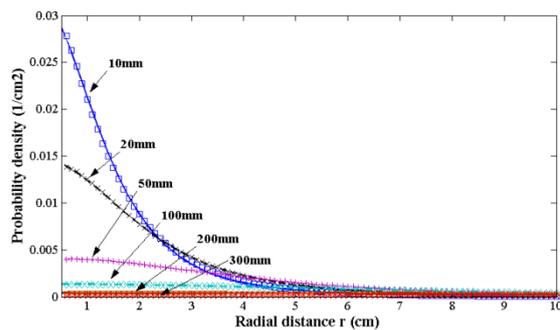

**Fig.8.** MCNPX simulated PScFs for polyethylene with thickness 20mm and different sample-to-detector distances.

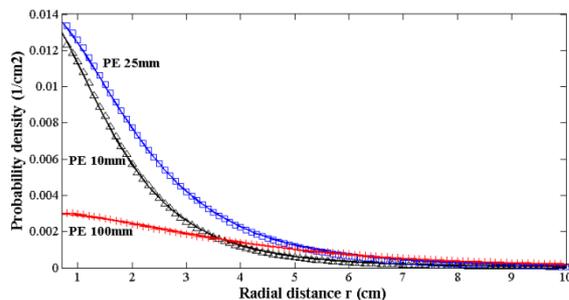

**Fig.9.** MCNPX simulated PScFs for PE for sample-to-detector distance 20mm and different thicknesses.

For the mathematical application of the PScF, Gaussian functions and isotropically scattered functions were used to fit the PScF funtions for thermal neutrons [10,11]. In this work, the simulated PScFs were well described by the following function with a very good agreement:

$$PScF(d,r) = \frac{S(d,t)}{(d^2 + r^2)^2} \quad (4)$$

where $t$ is the thickness of the sample, $d$ the sample-to-detector distance, $r$ the radial distance and $S(d,t)$ a polynomial expression of $t$ and $d$. The analytical functions of PScFs for polyethylene with different thicknesses and different sample-to-detector distances were also shown in Fig. 8 and Fig. 9. The results show good corresponding to the simulated data with the deviations less than 3%.

**3.2 Correction procedure**

The correction procedure was performed as follows:

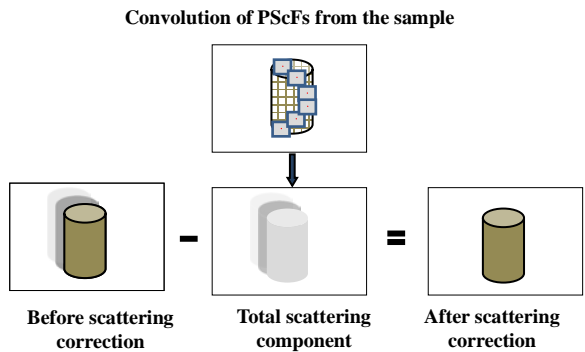

**Fig.10.** Procedure of scattering corrections using PScF

1. Deriving the thickness distributions of the sample from radiography images considering the neutron spectrum effects initially;

2. Calculating the total scattering component $I_s$ by a convolution of PScFs for each neutron beam in sample area based on PScF values;

3. Obtaining the corrected radiography image by a subtraction of initial image and scattering component $I_s$, as shown in Figure. 10

The whole scattering correction algorithm was implemented in a MATLAB code.

## 3.3 Experiments and results

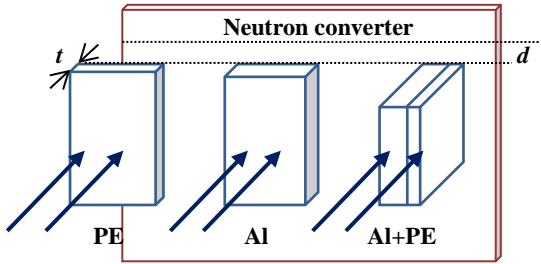

**Fig.11.** Experiment diagram of sample group 2. Three samples are used: PE ($t$=25mm), Al ($t$=40mm), a composition of PE+ Al ($t$=100mm). The sample-to-detector distance $d$ is changeable.

Three bricks made of polyethylene and aluminum were used for scattering corrections. The first is a 152×100×25mm³ polyethylene brick, the second a 152×100×40mm³ aluminum brick and the third a composition of PE and Al with thickness 100mm, as shown in Figure 11. The $d$ from back side of the samples to the fast neutron detector was: 10, 20, 50, 100, 200 and 300mm respectively.

The preprocessed radiographic images of sample group 2 are shown in Figure 12, Fig. 13 and Fig. 14 (a). It can be seen that the radiographies and related intensity profiles have sensitive responses with $d$. Fig. 12, Fig. 13 and Fig. 14 (b) show that scattered neutrons will make an extra contribution of 29% for PE and 53% for Al in the intensity profiles when $d$ is small, which would yield a big error in the information estimation of sample from exponential attenuation law. The smaller $d$ is, the larger the deviation will be. When $d$ is larger than 300mm, the intensity profiles will be convergent and the scatterings from the object can be neglected.

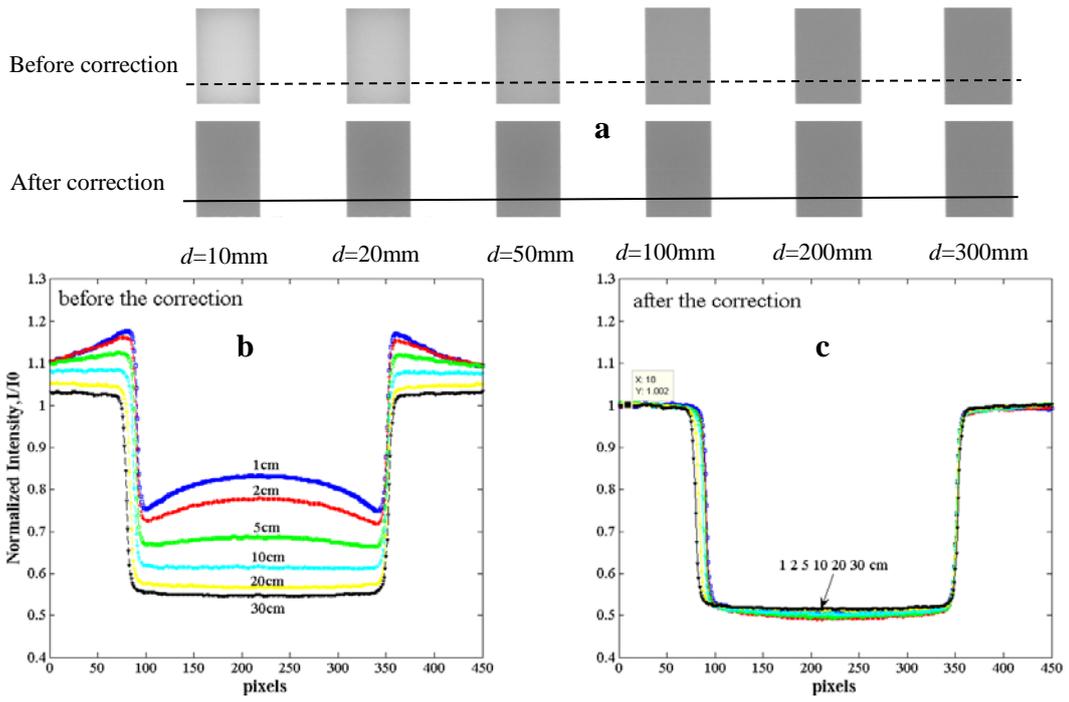

**Fig. 12** (a) Uncorrected and corrected radiographic images of Al for different d. The intensity profiles of the dashed line for (b) the uncorrected radiographies and (c) for the corrected radiographies.

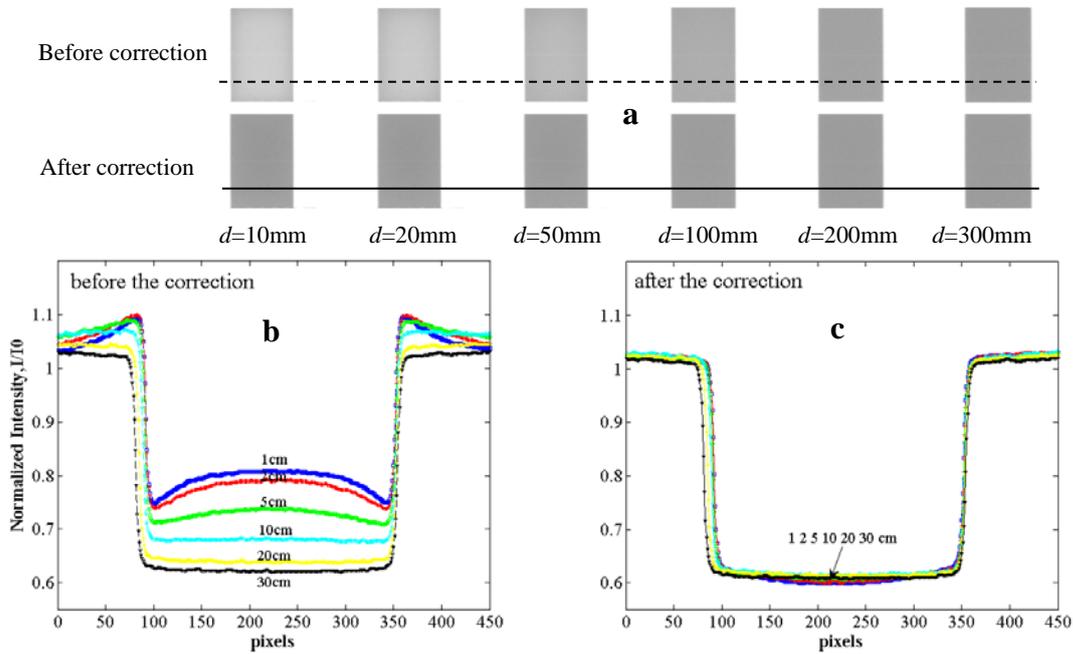

**Fig. 13** (a) Uncorrected and corrected radiographic images of PE for different d. The intensity profiles of the dashed line for (b) the uncorrected radiographies and (c) for the corrected radiographies.

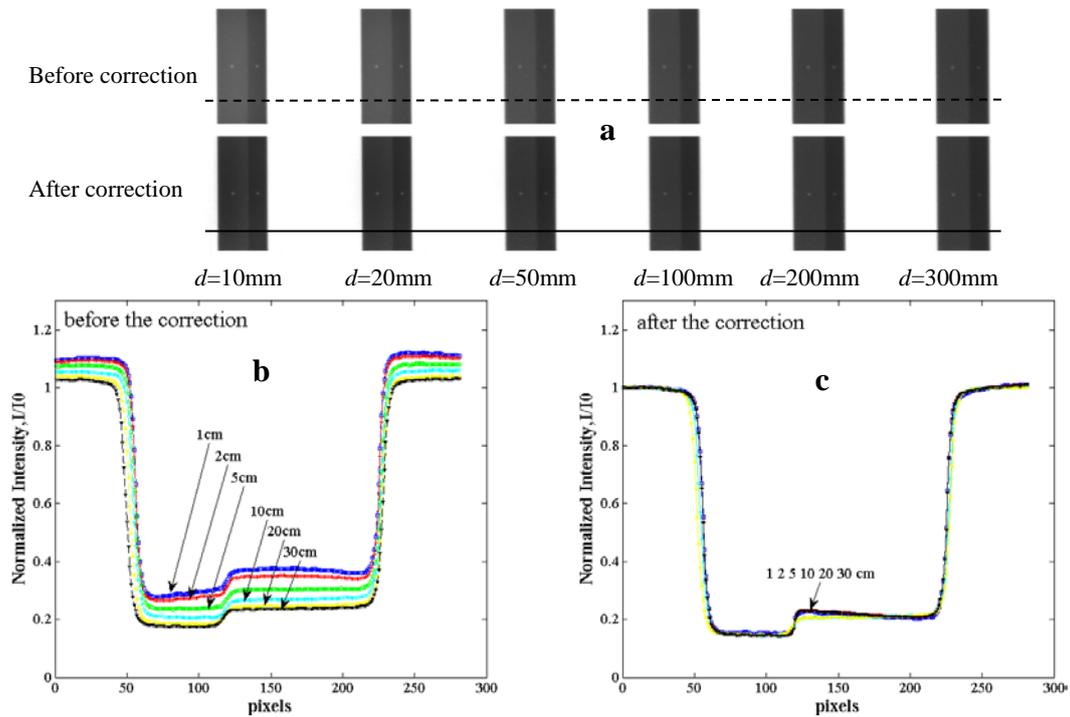

**Fig. 14** (a) Uncorrected and corrected radiographic images of PE+Al for different *d*. The intensity profiles of the dashed line for (b) the uncorrected radiographies and (c) for the corrected radiographies.

It can be seen in Fig. 12, Fig. 13 and Fig. 14 that after the scattering correction, the artifacts caused by scattered neutrons disappear. The normalized intensity profiles are now independent of the sample-to-detector distance $d$. A maximum improvement of 63% for Al, 34% for PE and 65% for PE+Al were achieved after applying the PScF scattering corrections. As an examination of quantitative analysis results, corrected images of 4cm-Al were used to derive the thickness distribution of different $d$ before and after the scattering corrections, with and without considering the neutron spectrum effects. The results are shown in Figure 15.

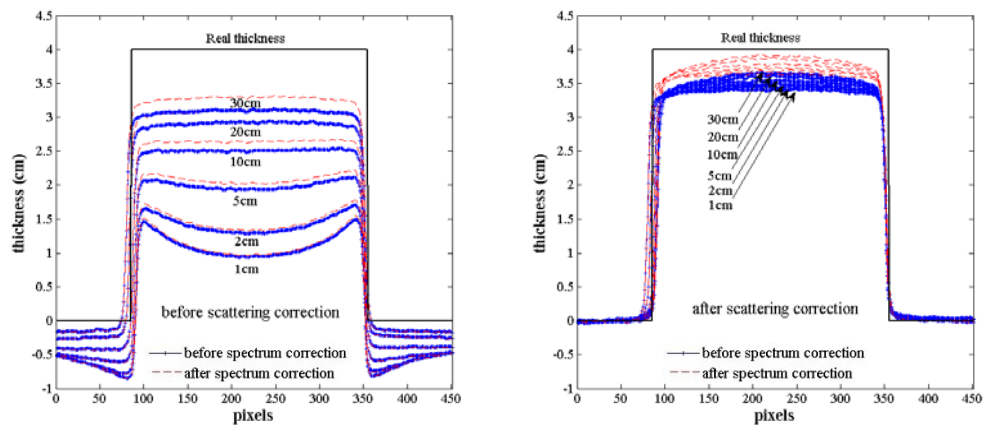

**Fig.15** Thickness distributions of 4cm Al derived from the intensity profiles for different $d$. Left: Before scattering corrections. Right: After scattering corrections. Bold lines: Consider the energy spectrum. Dashed lines: Not consider the energy spectrum.

It can be seen that the use of scattering corrections made a good improvement of the quantitative evaluation from the radiography images. Before the scattering correction, deviations from the obtained images are 33-75% even considering the spectrum effects. After the scattering corrections, deviations reduce to 8-15% for different $d$. Consider the energy spectrum of NECTAR, these deviations have a further reduction with an additional image improvement of about 5.5%.

## 4. Conclusion

For quantitative analysis and usage of fast neutron radiography in NECTAR, incident neutron spectrum and scattered neutrons from the sample are verified and corrected. Relationships of neutron transition intensities and penetration depths based on NECTAR spectrum were calculated and fitted for corrections. PScFs of fast neutrons for different materials were simulated and then fitted by analytical representations for further evaluations. A scattering correction algorithm was coded to evaluate the scattering components of the radiographic images. After eliminating the scattering components, the corrected images show identical and accuracy results. For verification, intensities of 4cm-Al before and after both corrections were compared and an improvement of 31-65% was obtained for different sample-to-detector distances.

## Acknowledgement

This work has been supported by the Academic Exchange Fund of Peking University Graduate School.

## References

[1] TANG Bin, ZHANG Song-Bao. Nuclear Electronics & Detection Technology, 2003, 23(2):176 (in Chinese)

[2] Richards W J et al. Applied Radiation and Isotopes, 2004, 61:551

[3] TANG Bin. Nuclear Electronics & Detection Technology, 2004, 24(4):387 (in Chinese)

[4] NALCIOGLU O, LOU R Y. 1979, PHYS. MED. BIOL, 24(2):330

[5] Incident neutron data, ENDF/B-VII

[6] Dou Hai-Feng, Tang Bin. CPC(HEP & NP), 2011, 35(5):483 (in Chinese)

[7] Chen Liang et al. Nuclear Physics Review, 2006, 23(3):310 (in Chinese)

[8] Zhang Fa-Qiang et al. ACTA PHYSICA SINICA, 2007, 56(6):3577 (in Chinese)

[9] Yoshii Koji et al. 1996, Nucl. Instrum. Methods A, 377:76

[10] Hassanein R et al. 2005, Nucl. Instrum. Methods A, 542:353

[11] Kardjilova N et al. 2005, Nucl. Instrum. Methods A, 542:336

[12] Bücherl T et al. 2011, Nucl. Instrum. Methods A, 651:86

[13] Guo J et al. 2009, Nucl. Instrum. Methods A, 605:69

[14] Hecht E. Optik. Bonn Munich, Addison-Wesley Publishing Company Inc., 1989

[15] http://www.lrz.de/services/compute/linux-cluster/doc/860-0237-001.pdf

[16] Breitkreutz Harald, Spektrale Charakterisierung des Therapiestrahls am FRM II (Ph. D. thesis). Munich: Fakultätfür Physik, Technische Universität München, FRM II, 2007 (in German)